# An Efficient Hybrid Localization Technique in Wireless Sensor Networks

Deepali Virmani[1], Satbir Jain[2]

[1]Department of CSE
BPIT, GGSIPU
Delhi, India.
deepalivirmani@gmail.com

[2]Department of CSE
NSIT, Dwarka
Delhi, India.
jain_satbir@yahoo.co.in

*Abstract—* Sensor nodes are low cost, low power devices that are used to collect physical data and monitor environmental conditions from remote locations. Wireless Sensor Networks (WSN's) are collection of sensor nodes, coordinating among themselves to perform a particular task. Localization is defined as the deployment of the sensor nodes at known locations in the network. Localization techniques are classified as Centralized and Distributed. MDS-Map and SDP are some of the centralized algorithms while Diffusion, Gradient, APIT, Bounding Box, Relaxation-Based and Coordinate System Stitching come under Distributed algorithms. In this paper, we propose a new hybrid localization technique, which combines the advantages of the centralized and distributed algorithms and overcomes some of the drawbacks of the existing techniques. Simulations done with J-Sim prove advantage of the proposed scheme in terms of localization error calculated by varying the sink nodes, increasing node density and increasing communication range.

## I. INTRODUCTION

Sensors are low-cost, low power devices with limited sensing, computation, and wireless communication capabilities [1]. A sensor consists of a sensing unit, a data processing unit, a communicating unit and a power supply unit. The power supply unit contains a battery that provides a limited amount of energy to the node.

A sensor network can be described as a collection of sensor nodes which co-ordinate to perform some specific action [2]. The positions of the sensor nodes may not be pre-defined. In case of changes such as the addition of new nodes, failure of nodes etc, the network efficiency should not be affected [3]. The region in which the nodes are deployed may not have any infrastructure. In this case, the nodes are responsible for their connectivity. Also, since the nodes have a limited amount of power backup, their energy should be optimally utilized so as to prolong the lifetime of the network. They have a variety of applications which include Military applications (monitoring forces, battlefield surveillance, battle damage assessment etc.), Environmental applications (forest fire detection, flood detection etc.), Health applications (tracking and monitoring doctors and patients inside a hospital, drug administration etc.), Home applications (home automation, smart environment etc.) and other Commercial applications (environmental control in office buildings, detecting and monitoring car thefts, vehicle tracking and detection etc.) [4].

Localization is the estimation of the positions of the sensor nodes in the network. It reduces power consumption and number of collisions, providing better accuracy. Localization techniques are classified as Centralized and Distributed. Centralized algorithms consist of a base station (sink node), to which all the sensor nodes transmit the collected information. After analysis of the received information, the computed locations are transported back into the network by the base station. As the network grows it increases the stress on the nodes near the base station. To prevent this multiple base stations may be used. Distributed algorithms, on the other hand, contain no such base station. Instead there exists inter-node communication to determine the locations independently.

## II. LITERATURE SURVEY

[5] Focuses on the concept of sensor networks and its architecture. It talks about its applications in detail. It also describes the factors influencing the sensor network design such as fault tolerance, scalability, production cost, hardware constraint, power consumption and environment.

[6][7] Describes the localization classification on the basis of Area of deployment, Physical layer, Parameter, Look up table, Estimation technique, and Security and Localising entity.

[8][9][10] Covers the problem of calculating the distance between the sensors by using Received Signal Strength Indication (RSSI), Radio Hop Count, Time Difference of Arrival (TDoA), Angle of Arrival (AoA). Further, it talks about the various centralised and distributed algorithms for localisation. Semi-Definite Programming (SDP) and Multi-Dimensional Scaling (MDS-MAP) are the two centralized approaches. In SDP, on the basis of the constraints between the nodes we calculate Linear Matrix Inequalities (LMI's). The LMI's are combined to form a single semi definite

program which is solved to produce the bounding region for each node. MDS-MAP uses the Law of Cosines and Linear Algebra to calculate the relative positions of the nodes based upon the pair-wise distances between them Diffusion, Gradient, Bounding Box and APIT are the BEACON-BASED Distributed Algorithms. Relaxation-Based and Coordinate System Stitching are NON BEACON-BASED Distributed Algorithms [11] [12] [13]. Diffusion sets the position of the non-beacon nodes to the average of its neighbour's location i.e. at the centroid of the neighbour's position. Gradient uses the concept of Multilateration for localisation in which the beacon nodes help their neighbour nodes to estimate their positions. The neighbouring nodes further assist their neighbours to find their positions. In Bounding Box the node location is at the intersection of the various bounding regions of the beacon nodes. In APIT, Approximate Point in Triangle, a node forms some Beacon-Triangles (consisting of three beacon nodes) and then localises its position at the centroid of the intersection of these triangles. Relaxation- Based technique starts with the nodes estimating their initial positions using any of the above described algorithms, which are then refined using the neighbours of the node as beacons. In Coordinate System Stitching, the network is first divided into small overlapping regions and local maps are constructed for each region. The local maps are then merged to form the global map of the network.

[14] [15] Discusses about the Localization methods classification and comparison between all the algorithms.

All the above references describe the different Localization techniques, their advantages, disadvantages and comparison between them. This paper presents a new hybrid technique, combining the centralized and distributed algorithms.

## III. THE CURRENT SCENARIO

The techniques discussed so far have their own advantages and limitations depending upon whether they are centralized or distributed.

In centralized algorithms [12] [16], as shown in figure 1(i) no computations need to be performed by the nodes since they communicate through the sink node, which does the computations for them. Also, the locations obtained are more precise. But, this leads to an increase in pressure on the sink node and once the sink node fails, the complete network collapses. Also, as the number of nodes increase, the energy efficiency decreases, the time delay increases and the traffic congestion also increases.

Distributed algorithms [17] [18], on the other hand, reduces the traffic congestion, has good scalability and has less storage requirements shown in figure 1(ii). The computational burden is distributed equally among all the nodes. But, there is no concept of using shortest-path for the inter-node communication, which leads to a decrease in throughput.

To overcome the above mentioned drawbacks of both the techniques, a new hybrid technique is proposed described in figure 2.

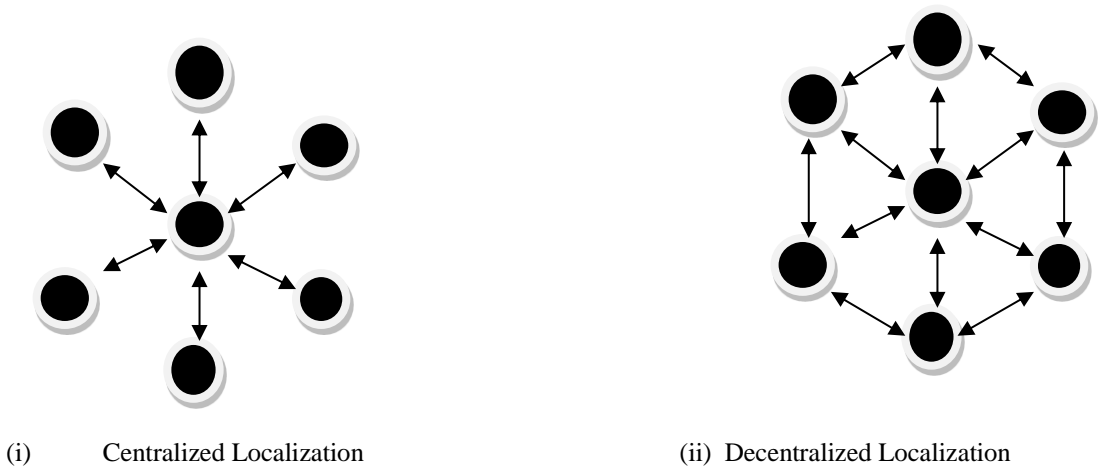

(i) Centralized Localization  (ii) Decentralized Localization

Fig1. Types of Localization

## IV. Hybrid Localization Technique

The techniques discussed earlier have some drawbacks as mentioned above. The proposed Hybrid technique (figure 2)

is a combination of the centralized and distributed localization algorithms. It includes centralized localization, a back-up node, a super-sink node for storing the database and distributed localization of all the sink nodes. It has the following functionalities:

*A.* Sink Node

*B.* Back-up Node
In each sub region a back-up node is created. It dynamically updates its information and works when the sink node fails.

The entire network is first divided into overlapping sub regions consisting of only one sink node. For example- The sink node along with its one-hop neighbour forms a sub-region. In the sub region so obtained a centralized algorithm like SDP (semi definite programming) or MDS-MAP (multi-dimensional scaling) is applied to localize the sensor nodes.

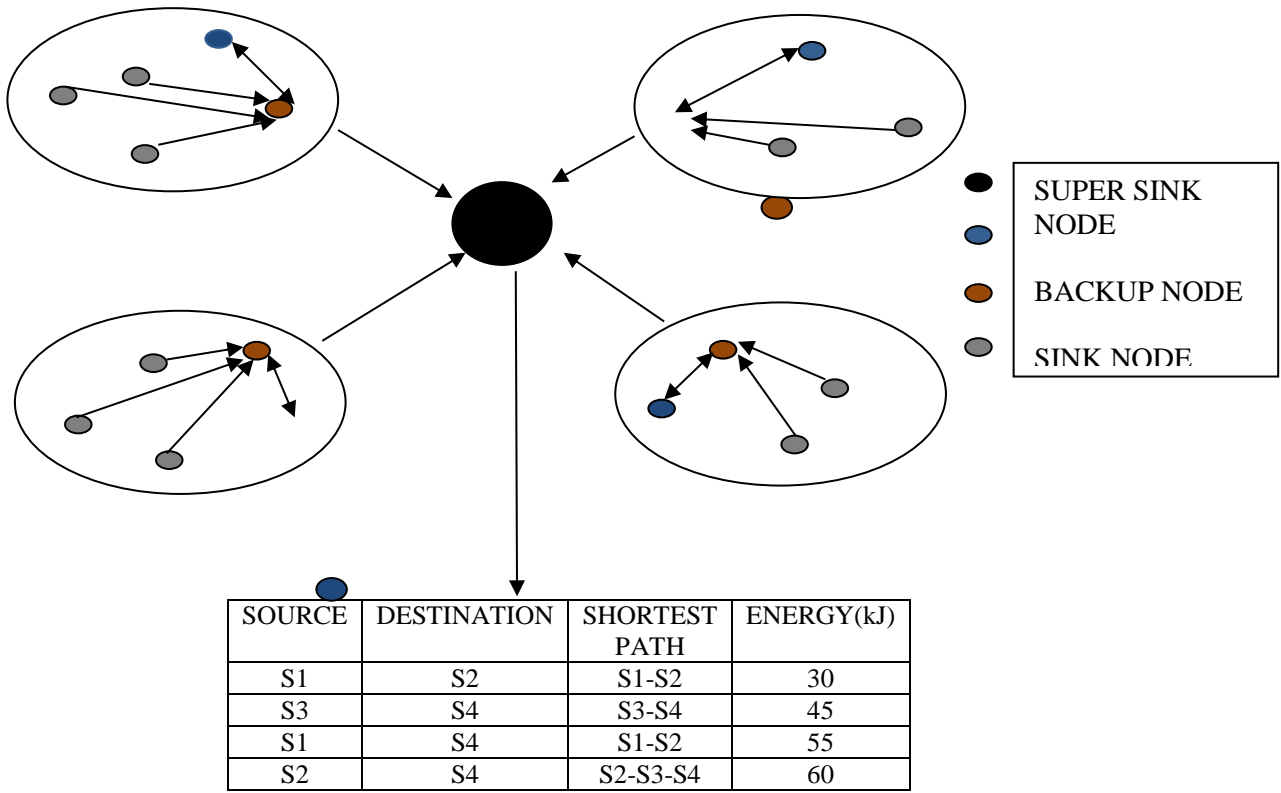

Fig 2. Hybrid Technique

*C.* Super-sink Node
The super sink node acts as a backbone and helps in the distributed localization of the sink nodes. It contains the information of all the sink nodes and generates a dynamic look up table, containing the shortest path that links the sink nodes. Whenever a new path is created, it automatically updates the table.

*D.* Sleep and Awake State
When a high density of beacons is deployed, only a certain percentage of the beacons involved in communication are activated. These sink nodes are in their AWAKE state. In case the energy of the beacon falls below the set threshold value, the sink node enters the SLEEP state. The super sink node may also force any of the sink nodes to enter the AWAKE or the SLEEP state as per the energy requirements and availability. It also controls the functionality of the back-up node. Initially, the backup node is in the SLEEP state. In case the energy of the sink node falls below the threshold value or it fails due to any reason, the super sink node sends an ALERT message to the backup node, which enters the AWAKE state in response.

The sensor nodes are low power devices. The amount of energy available for use is limited and is a constraint for the node's life. The concept of the AWAKE/SLEEP states of the sink nodes reduces the redundant use of energy since only the nodes involved in communication are active. Rest of the nodes are in the SLEEP state and hence save their energy. This leads to the efficient utilization of energy and power. The reduced energy and power consumption leads to an increase in bandwidth.

The backup node handles the condition of the failure of a sink node. It is a mirror image of the sink node and dynamically updates its information[18]. It is activated by the super sink node, through an ALERT message, in case the sink node fails or its energy level decreases beyond the threshold value. The backup node thus takes over causing no harm to the network. The dynamic updating of the lookup table by the super sink node leads to a fast access and reduces the time taken, thus increasing the throughput.

## V. HYBRID LOCALIZATION TECHNIQUE: PROPOSED ALGORITHM

Initially, we assume that every node in the network is present in the entire space with equal probability. We also assume that the network is fully connected, although our algorithm will also work in partitioned networks. The CU (Central Computing Unit) will execute the pseudo code shown in figure. 3. The CU processes each row in the log of every unknown node to obtain the constraints and updates the current estimate by intersecting it with the old position estimate. We will shortly explain what the addition and intersection symbols in the algorithm actually mean. If the beacon message is directly from a beacon, the constraint $C(x, y)$ imposed on the unknown node is given by a Gaussian normal distributed surface around the coordinates of the beacon.

Thus, the unknown node updates its position estimate by intersecting the old position estimate $P(x, y)$ with the constraint $C(x, y)$.

```
for (every unknown node)
    open the log file;
    initialize the position estimate P to the
    entire space;
  for (every row in the file)
    initialize constraint C to NULL;
    set pointer to mean1;
  while (!endof(row))
    read mean, stdev;
    compute new constraint N;
        C = C + N;
    increment pointer to
    point to the next mean;
  end while;
        P = P ∩ C ;
    end for;
        end for;
```

Fig 3 Pseudo code of hybrid technique

If the beacon message is from an unknown node, the CU has to process the cascade of distributions before intersecting with the old position estimate. The new constraint is calculated by adding the individual constraints. The sum of these constraints is similar to the convolution of all the individual distributions. Assume that we have a beacon at coordinates $(x_b, y_b)$ and two unknown nodes 1 and 2. Assume that, corresponding to the signal strength that node 1 receives from the beacon we have the function $f1(d)$ (e.g., a Gaussian with parameters $\mu 1$ and $\sigma 1$) and similarly, corresponding to the signal strength that node 2 receives from node 1 we have the function $f2(d)$. Then, the position estimate of node 1 is given by equation 1

$$Ep = f1\left(d(x,y), (x_b, y_b)\right) \forall (x,y) \qquad (1)$$

Where $d(x,y)$ is the Euclidean distance between points x, y. The position of node 2 is calculated using equation 2

$$Ep' = \frac{\int_{x1}^{x1} \int Ep(x,y)f2\left(d((x,y),(x1,y1))\right)dx1,dy1}{\int_{x2}^{x2}\int_{y2}^{y2}\int_{x1}^{x1}\int_{y1}^{y1} Ep(x1,y1)f2\left(d((x2,y2),(x1,y1))\right)dx1,dy1,dx2,dy2} \qquad (2)$$

Present position is evaluated by equation

$$P(x,y) = \frac{p(x,y) \times C(x,y)}{\iint_{-\infty}^{\infty} p(x,y) \times C(x,y)dx\,dy} \qquad (3)$$

## VI. COMPARISON SIMULATION AND PERFORMANCE EVALUATION

In this paper, to evaluate the performance of the proposed technique using J-Sim, we have set up the following simulation conditions:

(i) Unknown sensor nodes are deployed randomly in area of 100 x 100 m2.
(ii) The sensor nodes are distributed grid wise and the number of nodes is set to 85.
(iii) The radio range of sensor node (r) is set to 30.

In our simulations, we study several system-wide parameters that can affect localization error. In our work, we study several system-wide parameters that can affect localization error. These parameters are:

- Sink Node (SN): These are the nodes whose location is known and in our experiment, values are set to 4,9,16,25,30,49,100 as deployed grid wise.
- Communication Range (CR): This is the range or area or the propagation distance to rest of the nodes, and is varied from 20 to 100.
- Node Density (ND): These are total number of nodes in a network. In these simulations, unknown nodes are varied from 50 to 90.

Localization error (E) can be calculated by

$$E = \frac{\sqrt{(x'-x)^2+(y'-y)^2}}{CR} \quad (4)$$

Where (x', y') are the nodes estimated coordinates, (x, y) are the nodes real coordinates and CR is the communication range.

*A. Localization Error when Sink Node is varied:*

In this simulation, we analyse the effect of change of number of sink nodes on the localization error. Localization error is estimated on centralized, decentralized and the proposed hybrid technique by taking different values of sink nodes. Figure 4 show that estimation error decreases as the number of anchor nodes increases. Hybrid technique localization scheme shows constant decrease in localization error with increase in sink nodes. As it can also be observed that proposed hybrid technique has smaller location error in general with respect to different number of sink nodes.

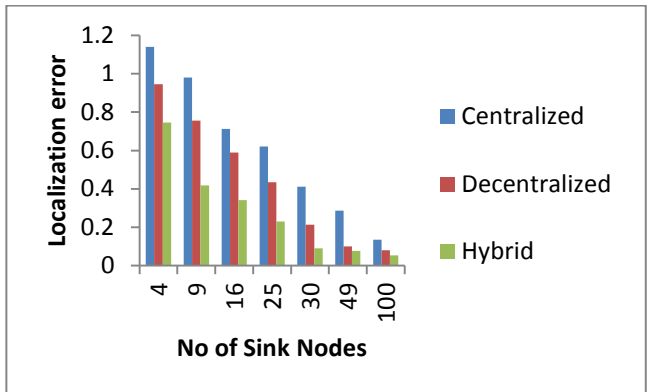

Fig 4 Localization error with respect to no of sink nodes

*B. Localization Error when communication range is varied:*

In Figure 5, where graph represents that with the increase in communication range, estimation error increases. This is due to the fact anchor propagation distance result in larger accumulated error. There is significant increase in error with the increase in anchor to node range or communication range (CR). Large numbers of sink nodes are desired for good estimation results. The cost of having such a large percentage of anchors is very high so instead of increasing anchor nodes, we can increase the anchor radio range to which beacons travel. Here we can observe that for proposed technique, gained estimation error is least.

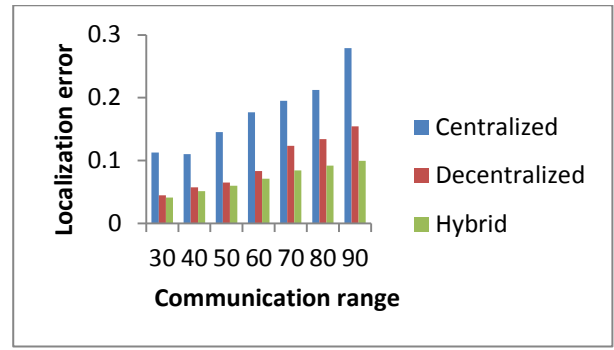

Fig 5 Localization error with respect to Communication range

*C. Localization Error when node density is varied:*

In this, we analyze the effect of node density on the localization error in a network area. Figure 6 given below shows the values of node density are varied from 50 to 90 and how change in node density affects the error for different schemes.

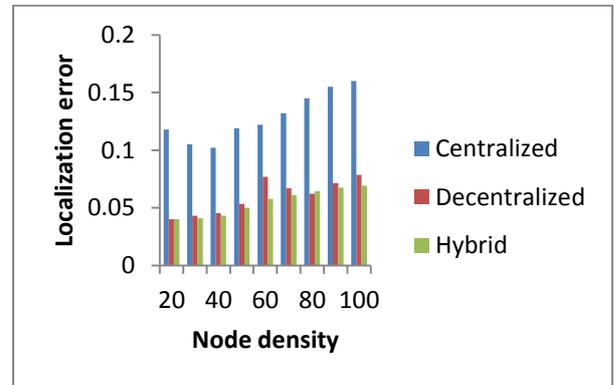

Fig 6 Localization error with respect to node density

VII Conclusion

Localization in Sensor Networks is a vast field of research. In this paper, we first discuss the various localization techniques available and then present a new hybrid localization technique that is a combination of centralized and distributed techniques. Simulation results are evaluated by varying different node parameters such as number of sink nodes, node density, and communication range. From the simulation curves results we can conclude that our distance based hybrid technique improves the positioning accuracy significantly. The hybrid technique has large estimation error; it remains independent of node density. In conclusion, hybrid technique demands no additional hardware to implement combination